\documentclass[twocolumn]{article}

\usepackage{amsmath,amssymb,amsfonts}
\usepackage{algorithmic}
\usepackage{graphicx}
\usepackage{textcomp}
\usepackage{wrapfig}
\usepackage{hyperref}
\hypersetup{hidelinks=true}

\usepackage[style=ieee, isbn=false, url=false, eprint=false]{biblatex}
\AtBeginBibliography{\footnotesize}

\graphicspath{{Figures/}}

\begin{document}

\renewcommand\footnotemark{}

\title{\vspace{-2cm}A Method to Achieve High Dynamic Range in a CMOS Image Sensor Using Interleaved Row Readout}

\author{Thomas Wocial$^{1}$, Konstantin D. Stefanov$^{2}$, William E. Martin$^{1}$, John R. Barnes$^{2}$, \and Hugh R.A. Jones$^{1}$
\thanks{Manuscript received 13/08/2022; revised 12/09/2022; accepted 14/09/2022. Date of publication 06/10/2022; date of current version 10/10/2022. TW, WM and HJ acknowledge support from STFC grants ST/W507490/1, ST/W508020/1, ST/P005667/1 and ST/R006598/1. This work was made possible through a Research, Enterprise and Scholarship Innovation grant from the Open University, which funded both KDS and JRB.}
\thanks{$^{1}$TW, WM and HJ are with the Centre for Astrophysics Research, University of Hertfordshire, College Lane, Hatfield AL10 9AB, UK (e-mail: t.wocial@herts.ac.uk).} 
\thanks{$^{2}$KDS and JRB are with the School of Physical Sciences, Open University, MK7 6AA, UK (e-mail: konstantin.stefanov@open.ac.uk).}
\thanks{Digital Object Identifier: \href{https://doi.org/10.1109/JSEN.2022.3211152}{10.1109/JSEN.2022.3211152}}}

\date{}
\maketitle

\begin{abstract}

We present a readout scheme for CMOS image sensors that can be used to achieve arbitrarily high dynamic range (HDR) in principle. The linear full well capacity (LFWC) in high signal regions was extended 50 times from 20 ke$^{-}$ to 984 ke$^{-}$ via an interlaced row-wise readout order, whilst the noise floor remained unchanged in low signal regions, resulting in a 34 dB increase in DR. The peak signal-to-noise ratio (PSNR) is increased in a continuous fashion from 43 dB to 60 dB. This was achieved by summing user-selected rows which were read out multiple times. Centroiding uncertainties were lowered when template-fitting a projected pattern, compared to the standard readout scheme. Example applications are aimed at scientific imaging due to the linearity and PSNR increase.
\end{abstract}

\section{Introduction}
In recent years, the CMOS image sensor (CIS) has seen increased adoption by the astronomical community, particularly in time-domain applications \cite{Pratlong2016}. First invented in 1993 \cite{Fossum1997}, some advantages of the active pixel architecture over charge-coupled devices (CCDs) include 1) ability for faster readout, 2) readout from region of interest (ROI), 3) low power, 4) low readout noise and 5) built-in anti-blooming. 

A variety of techniques exist to extend the dynamic range (DR) of CMOS image sensors. These have previously been classified into seven categories: 1) logarithmic pixel response, 2) combined linear and logarithmic response, 3) well capacity adjustment, 4) frequency based sensors, 5) time-to-saturation based sensors, 6) global control of integration and 7) local control over integration \cite{Yang1999}\cite{Spivak2009}. Other advances that result in an increased DR include improved dark current suppression, lower read noise and multi gain readout \cite{fowler5Mpixel100Frames2010}.

The 4/5T pixel architecture is widely employed in scientific imaging, but the selection of DR extension techniques is limited. For instance, dynamic ranges of 160 dB have been achieved in CIS using a 3T architecture with integrated charge compensation photodiode \cite{Li2016} and 141 dB (proposed) on 4T architecture with a lateral overflow integration capacitor but both suffer from non-linear pixel responses and signal-to-noise ratio (SNR) curves. DR extension techniques used for comparison to this work will apply to 4T pixels and therefore fall under: local control over integration and global control over integration. 

For the scientific requirement and operation of our CIS, key assumptions are made about the scene to be imaged: 1) there exists a desired minimum SNR of the observation, hence minimum integration time, 2) the scene is spatially and temporally static, 3) it is always desirable to achieve the maximum possible SNR for any local region. 

Self-reset pixels achieve HDR by locally controlling integration time. The pixel is based on the 4T architecture with additional circuitry to trigger a reset signal when the voltage at the sense node matches a reference corresponding to the full well capacity (FWC). By counting the triggers and sampling the residual voltage it is possible to reconstruct photosignals that would otherwise exceed the pixel FWC for a given integration whilst preserving linearity. Recent implementations include a 16$\times$16 pixel on 20 $\mu$m pitch at 121 dB DR \cite{Lenero-Bardallo2017} and 96$\times$128 pixel on 25 $\mu$m pitch at 125 dB DR \cite{Hirsch2019}. Fill factors are 13.1$\%$ and 10$\%$ respectively. A key advantage is the continuous SNR increase with photosignal, however the low fill factor, large pitch and small array sizes can be a challenge for use in scientific imaging.

Multiple exposure is a commonly employed technique to increase DR without the need for additional circuitry. Dual exposure involves acquiring two images at different integration times $T_{long}$ and $T_{short}$ with the DR extension equal to $T_{long} / T_{short}$ \cite{is__ts_conference_is_1993}\cite{yadid-pechtWideIntrasceneDynamic1997}. There exists an SNR dip in regions corresponding to saturation in the short exposure, as the photosignal here is only sampled for a fraction of the total integration time. The use of multiple shorter integrations can lower the resulting SNR dip \cite{sasakiWideDynamicRangeCMOSImage2007b}.

Non-destructive readout (NDR) operates by sampling the signal many times during the integration time without affecting the built up photocharge. The CIS is read with up-the-ramp sampling, as widely used on IR photodiode arrays for DR increase \cite{kachatkouDynamicRangeEnhancement2008}\cite{simsCMOSVisibleImage2018} and for cosmic ray rejection \cite{offenbergValidationRampSampling2001}.

Coded rolling shutter \cite{guCodedRollingShutter2010} works by spatially varying exposure per row, achieving a DR increase up to the ratio of longest and shortest exposure times. Image reconstruction is needed as vertical spatial information may be lost, however, due to the encoded temporal information, high speed video and optical flow can be extracted \cite{liuEfficientSpaceTimeSampling2014a}\cite{choSingleshotHighDynamic2014}. Interleaved multiple gain readout has been used to achieve a DR of 120 dB \cite{dupontDualcoreHighlyProgrammable2016}. Work in this area is limited by commercial CIS devices often being addressable row-wise only, pixel-wise coded exposure could be achieved with full X-Y addressability \cite{liuEfficientSpaceTimeSampling2014a}.

Pixel-wise control of integration times builds on row-wise coded exposure by the addition of control over an additional spatial dimension. Recent advances in 3D stacked CIS devices haven enabled a pixel-parallel architecture, typically based on Cu-Cu interconnects between the sensing and logic layers \cite{oikeEvolutionImageSensor2022}. A similar method is to control exposure for a block of pixels, as shown in \cite{hirata1inch17Mpixel1000fps2021}\cite{Peizerat2015}. In \cite{ikeno6mm127dBDynamic2022} a 512$\times$512 array with 4.6 $\mu$m pitch on a 3D-IC achieves 127 dB DR by combining dual conversion gain with time-to-saturation detection. Key advantages are the high spatial fidelity achieved with a pixel-parallel readout, however there is non-continuity in the SNR response due to the three photosensing regimes.  

In this paper a row-wise coded exposure scheme is proposed which varies exposure locally depending on the illumination, with multiple sub-exposures occurring in said regions to exceed the SNR limit imposed by the FWC of the pixel. A hardware demonstration on a Teledyne e2v (Te2v) SIRIUS CIS115 \cite{e2v} results in extended DR and peak signal-to-noise ratio (PSNR). This is achieved thanks to the ability to randomly address and read out rows. A future implementation of this technique is for actively controlled spectroscopy \cite{Jones2021}, as both high signal calibration spectral lines and science spectral lines fall on the focal plane \cite{Li2008}\cite{2010exop.book..111T}. Other applications may include multi-ROI imaging at varying sub-exposures and other areas, where high DR and SNR images are of interest. The ability for CMOS image sensors to operate in such way has been recognised by others \cite{Knapp2020}\cite{Greffe}, but to our knowledge has not been implemented in scientific imaging applications. 

\section{Dynamic range}
A scene's dynamic range is given by the upper and lower limits of its luminance range. It is desirable for a CIS to have a high dynamic range as this means it can faithfully quantify both the high and low signal regions in the scene. For a sensor with a linear full well capacity of Q$_{well}$ (e$^{-}$), exposure time $t$ (s), average dark current $I_{dc}$ (e$^{-}/$s) and read noise variance $\sigma_{read}^{2}$ (e$^{2}$), the DR is the ratio of maximum ($I_{max}$) and minimum ($I_{min}$) detectable photocurrents as in equation \ref{DR basic}. The maximum detectable signal is limited by Q$_{well}$ minus the dark signal (which is subtracted from a reference image taken in darkness), whilst the noise floor (assuming correlated double sampling and fixed pattern noise correction) is determined by the read noise and dark current signal noise.

\begin{equation}\label{DR basic}
    DR\ =\ 20\ log_{10} \frac{I_{max}}{I_{min}} = 20\ log_{10} \frac{Q_{well} - I_{dc}t}{\sqrt{\sigma_{read}^{2} + {I}_{dc}t}}
\end{equation}

The readout noise determines the noise floor of a CIS. This value is almost independent of temperature and integration time. It is comprised of transistor noise from the source follower, amplifier and analogue-to-digital converter (ADC) noise. Quantization noise can contribute to read noise if the signal is sampled with low resolution. To determine the readout noise we calculate the standard deviation of each pixel using a set of bias frames \cite{Starkey2016}. The input-referred noise is often given in equivalent noise charge, expressed in e$^{-}$ RMS.

Dark current occurs as a result of thermal excitation of electron-hole pairs in silicon. The sensor used in this paper, the CIS115, has a dark current of 20 e$^{-}$/pix/s at 293 K, which halves for every temperature reduction of 5.5 K \cite{e2v}. If the noise floor is dominated by dark current, using shorter exposures or cooling the CIS can increase the DR.

\section{Method}
\subsection{CIS115}

The sensor used is a CIS115 from Te2v, featuring an array of 2000 $\times$ 1504 pixels on 7 $\mu$m pitch \cite{e2v} (see Table \ref{tab:Specs} for measured BSI variant specifications). It employs the 4T pixel architecture and is fabricated using a 0.18 $\mu$m CIS process. The model used in this paper is a front side illuminated variant. A back side illuminated variant has been adopted for the JANUS instrument on JUICE \cite{Soman2014}\cite{Soman2015}. The sensor is divided into four blocks of 376 columns which are read out simultaneously and in parallel. Each pixel transfers the reset and signal levels to a storage buffer that allows for correlated double sampling (CDS) to be performed. The CIS115 operates in rolling shutter mode meaning integration time is simultaneous for all pixels in a row. All control signals are generated externally.

\begin{table}[h]
\centering
\begin{tabular}{llll}
\multicolumn{1}{c}{}  & Unit & \multicolumn{1}{c}{CIS115}  \\ \hline
Active rows           &                    & 2000                                 \\
Active columns        &                    & 1504                                  \\
Pixel size            & $\mu$m             & 7.0                                  \\
Non-linearity         & $\pm\%$            & 3                                  \\
Mean read noise       & e$^{-}$            & 5                                \\
Peak linear charge    & e$^{-}$/pix        & 27,000                              \\
Saturation charge     & e$^{-}$/pix        & 33,000                              \\
Dynamic range         & dB                 & 74.6                                \\
Dark current     & e$^{-}$/pix/s      & 20 (at 293 K)                      \\
DSNU                  & e$^{-}$/pix/s      & 12 (at 293 K)                       \\
\end{tabular}
\caption{Specifications for the CIS115 \cite{e2v}}
\label{tab:Specs}
\end{table}

\subsection{Row-wise HDR readout scheme}

The scheme demonstrated in this paper seeks to address key traits for scientific imaging: pixel linearity and continuous SNR increase with photosignal. For a maximum per pixel photocurrent per row $I(M)$, a total integration time $t$, with row-wise control over exposure, the scheme is as follows: If the FWC is reached or exceeded in $t$, perform $N$ sub-exposures with a read and reset sample such that equation \ref{N row} is satisfied. N rows are summed (stacked) in software.

\begin{equation}\label{N row}
    N = \lceil I(M) \times t / FWC\rceil
\end{equation}

By doing so, the noise floor is kept at a minimum on a per-row basis and the PSNR is increased by sampling the maximum detectable photosignal multiple times. We implement a simplified version of this scheme with one region that can reconstruct photosignals 50 times greater than the FWC per integration.

The CIS115 is mounted on a control PCB, with sensor interfacing achieved via a National Instruments PXIe-7856R FPGA card. A custom LabVIEW GUI was developed at the Open University. In standard configuration the minimum row readout time, $T_{row}$, is 412.5 $\mu$s so the readout time for the whole array once is 825 ms. During readout the analogue sensor outputs are digitised by four 16-bit ADCs in the PXIe-7856R card with the digital values stored as signed 32 bit integers. In this demonstration HDR is achieved by sorting the rows ($M_{Total}$) into two groups: rows read out once ($M_{Once}$) and rows read out $N$ additional times per integration ($M_{HDR}$). Integration time for rows read once $T_{Once}$ (equation \ref{T once}) is therefore $N$ times larger than those read multiple times $T_{HDR}$ (equation \ref{T multiple}). The signals from rows read multiple times are then summed to achieve the same total integration time with increased equivalent full well capacity.

\begin{equation}\label{T once}
    T_{Once} = T_{row} \times (M_{Total}+N\times M_{HDR})
\end{equation}
\begin{equation}\label{T multiple}
    T_{HDR} = T_{row} \times (M_{Total}/N+M_{HDR})
\end{equation}
\begin{figure}[h]
    \centering
    \includegraphics[width=0.48\textwidth]{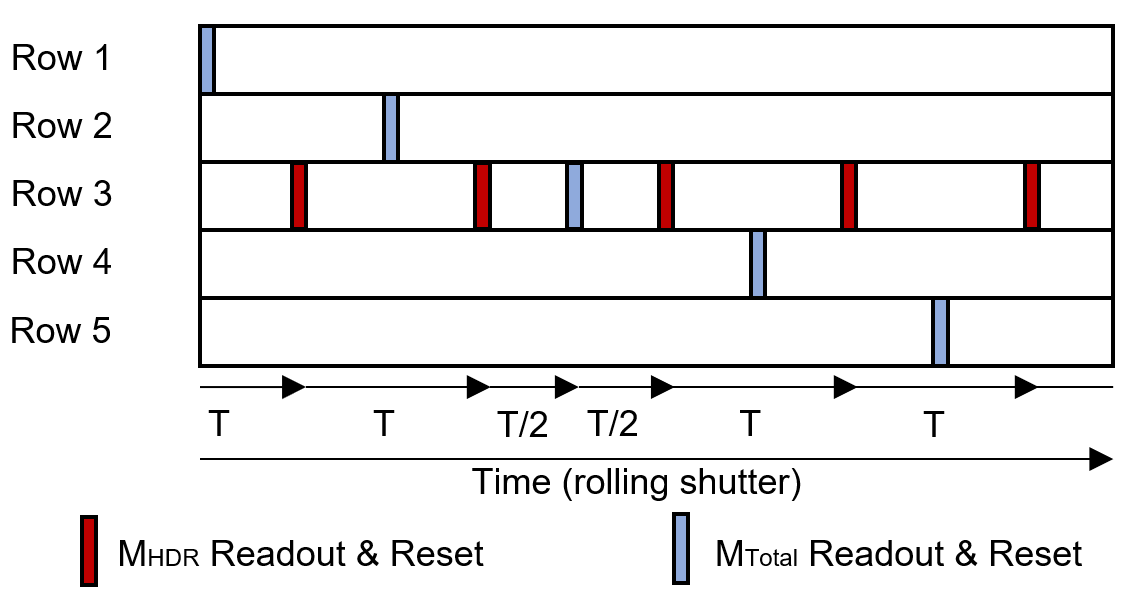}
    \caption{Simplified row-wise timing diagram in rolling shutter operation. In this example for $N$=5, row three is read out six times in one integration whilst the other four rows just once. Integration occurs in the white regions with readout and reset samples in the blue for the standard readout and in red for the HDR readout.} 
    \label{fig:timing}
\end{figure}

Rows read out multiple times have to be interleaved with rows read once to keep constant timing, shown in a simplified timing diagram in Figure \ref{fig:timing}. Here the read noise in row three increases by a factor of $\sqrt{5+1}$ and effective full well capacity by a factor of 5. A readout is always accompanied by a photodiode reset. Each block is the sum of the CDS time (10 $\mu$s) and the row readout time for a total of 412.5 $\mu$s (not to scale). A delay timer can also be added, where no readout occurs. The total number of row readouts is $M_{Total}+N\times M_{HDR}$. The order in which rows are read is determined by $M_{total}$ and $N$. In the HDR readout configuration used these are 2000 and 50, respectively. If $M_{total}$ does not divide by $N$ to give an integer, then rows in $M_{HDR}$ will not have constant timing due to rounding to the nearest row. The 40 rows used for HDR, $M_{HDR}$, are from rows 1495 to 1534 in order to image the brightest centroid at row 1512. Each additional row read and reset sample increases the integration time by 412.5 $\mu$s. As 4000 row read and reset samples occur, the integration time is 1650 ms. This constrains use cases towards lower peak luminance HDR scenes, though a CIS with a shorter sample time would overcome this.

\begin{figure}[h]
    \centering
    \includegraphics[width=0.5\textwidth]{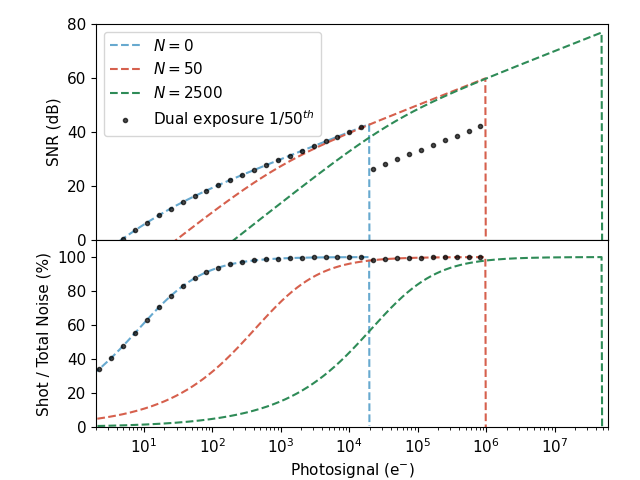}
    \caption{Noise performance at varying photosignals for the HDR extension scheme described in this paper. $N$ is the number of additional row signal and reset samples per integration.}
    \label{fig:PSNR}
\end{figure}

The last point is demonstrated with simulated performance in Figure \ref{fig:PSNR}. The term $N$ represents the number of additional row reads (hence stacked additional sub-frames) in a given region. With appropriate selection of row readouts and resets summed in a given integration period, the PSNR and maximum DR can be increased arbitrarily high with the photosignal (as in $N=2500$) whilst simultaneously retaining improved noise characteristics in low signal regions ($N=0$). The noise model is simplified, comprising of read and shot noise only. Read noise and LFWC are the values found for our device at 4.1 e$^{-}$ and 19.7 ke$^{-}$ respectively. For comparison, a dual exposure scheme with a 1:50 exposure ratio is shown by black markers. Non-destructive readout and single read row-wise coded exposure exhibit equivalent SNR profiles. DR is extended by 34 dB as with our method for N=50, although there is a significant drop in SNR from 42.9 dB to 25.8 dB when the photosignal exceeds the FWC, with the PSNR remaining unchanged. In our method SNR drops by only 0.2 dB and continues to rise to 59.9 dB over the same region.
  
\begin{figure*}[h]
    \centering
    \includegraphics[width=0.65\textwidth]{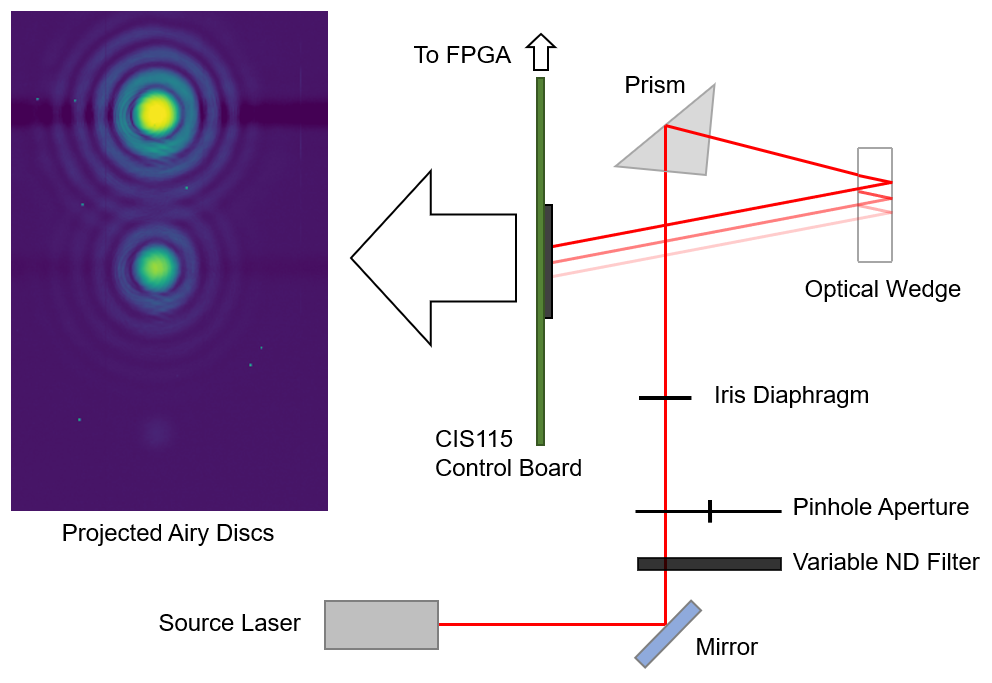}
    \caption{Optical setup used. Laser source is a Thorlabs CPS532-C2, ND filter a Thorlabs NDC-50C-4, mirror, Thorlabs PHWM16 pinhole, 45 degree prism, a Thorlabs  WW41050-A optical wedge and the CIS115 board. The optics are used to project the three Airy Discs shown in the top left}
    \label{fig:Setup}
\end{figure*}

Once integration is complete the image is saved as a 32 bit binary file with original sensor dimensions. When read with this HDR scheme, the effective full well capacity is increased by a factor of $N$ and read noise increased by $\sqrt{N+1}$ for rows in $M_{HDR}$, with $M_{Once}$ retaining original full well capacity and read noise. DR for the whole array is extended by a factor of $20 \log_{10}(N)$ dB.

As a result the DR of an imaged object is dependant on the sensor rows on which it appears. In this work, the limited readout bandwidth resulted in long integration times so dark current noise contributions was comparable to read noise. This would ideally be reduced with higher readout rates. Timer control or reading dummy rows allow for more flexible readout orders. In addition, only two regions are used in this paper, though schemes may have regions with multiple varying cadences.

\subsection{Optics}
A dynamic range measurement can be performed using a flat field with either variable source intensity or integration time. This is achieved by measuring the linear FWC and read noise. This method is simple to perform, however does not allow for simultaneous noise analysis in low signal regions. By projecting a HDR pattern onto the CIS, signal analysis can be performed for a continuous range of intensities allowing for quantification of artefacts introduced by the HDR extension technique. 

Analysis is performed using the three reflected Airy disc patterns projected onto the focal plane (Figure \ref{fig:Setup}). First, a collimated 532 nm, 0.9 mW laser (Thorlabs CPS532-C2), is attenuated using an adjustable neutral density filter. A 300 $\mu$m pinhole aperture (Thorlabs PHWM16) produces the Airy disc with an adjustable iris used to spatially filter the beam to mitigate unwanted reflections. The beam profile is then picked off by right angled prism, to an optical wedge opposite the CIS115, at angle of 30\textdegree\ (Thorlabs WW41050-A). Fine adjustment of the angles of the prism and wedge allow for spatial position of the pattern. The wedge can also be rotated. In this case the wedge is rotated such that the three reflections lay on an axis parallel to the columns of the CIS115, as in Figure \ref{fig:Setup}. The spatial filter was adjusted to block maxima greater than four. This helps mitigate against specular reflection from flat surfaces propagating through the optical path and stops the outer orders from the first Airy pattern swamping the signal from the second and third pattern. The laser was kept on for at least 10 minutes before taking measurements to mitigate heating effects.

\begin{equation}\label{airy}
    I(r_{p}) = I_{0} \left(\frac{2 J_{1}(r_{p}ka\mu/h)}{r_{p}ka\mu/h} \right)^{2}
\end{equation}

In pixel units, the radial variation of intensity from the centroid on the image plane is given by equation \ref{airy}, where $r_{p}$ is the distance to the centroid in pixels, $I_{0}$ the intensity normalisation constant to be found, $J_{1}$ the first order Bessel function of the first kind, $k$ the wavenumber, $a$ pinhole radius ($150 \mu m$), $\mu$ pixel size (7.0 $\mu m$) and $h$ the path length from pinhole to image plane. The small angle approximation is employed to convert to pixel units. Using a standard readout scheme and ensuring the Airy disk remained in the linear region, a line profile was taken and used to fit the Airy profile above. The value of $ka\mu/h$ was found to be equal to $3.89\pm0.01 \times 10^{-2}$ pix$^{-1}$. As other constants are known, $h$ solves to be 318$\pm$1 mm, exceeding the criterion for far-field diffraction of $D^2/\lambda$. Setting the LHS of equation \ref{airy} as $I_{0}/2$ and solving for $r_{p}$ gives a FWHM of  $83.1\pm0.2$ pixels, demonstrated later in Figure \ref{fig:reflection 1}.

\subsection{Artefacts}

Image corrections are performed. The top left of Figure \ref{fig:Setup} is a 825 ms integration with the three Airy discs visible. A 825 ms dark integration has been subtracted for correction of DSNU and FPN. Ten hot pixels are also visible in this image. Correction is applied using a bilinear interpolation. Four bad columns are replaced using linear interpolation. The final feature is a signal dependant baseline shift from the expected value. That is, rows which fall on high signal regions deviate below the baseline value in low signal regions. This feature is visible in Figure \ref{fig:Setup} (top left) as the darker blue horizontal bands which occur either side of the maxima of the first and second Airy disc. This is due to an offset of the ground return to the source follower column bias. This feature has been reduced significantly in some applications by creating dark reference pixels for correction

It was found that the baseline deviation was proportional to the sum of all pixel values in a row. With the CIS115 operating in HDR mode and the three Airy discs projected onto the focal plane, the median background level for a row is plotted against the sum of all pixels in the row ($\Sigma_{Row}$) in Figure \ref{fig:Setup}. Using a least squares linear fit, the expected row-wide deviation is equal to $-1.339 \times 10^{-5}$ ADU per $\Sigma_{Row}$. The light frames are then corrected using this relationship by multiplying with the row sum and adding to each pixel before linearity correction and conversion from ADU to e$^{-}$.

\subsection{Sensor characterisation}

\begin{figure}[h]
    \centering
    \includegraphics[width=0.5\textwidth]{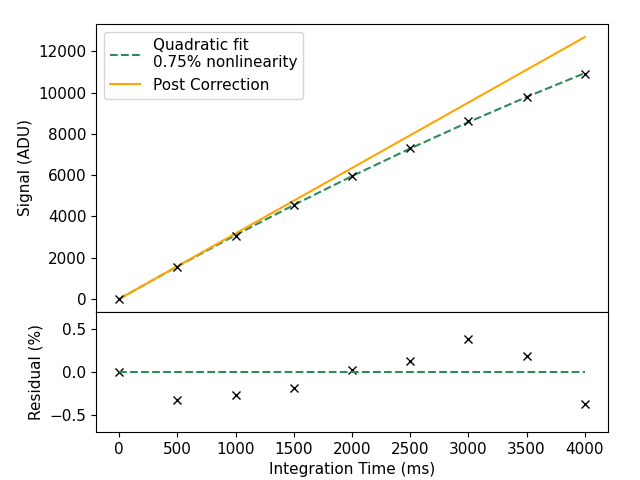}
    \caption{Mean signal of a 200 by 200 pixel area against increasing integration times. Residual plot is the deviation from the linear regression as a percentage.}
    \label{fig:Linearity}
\end{figure}

\subsubsection{System gain}
The system gain is defined as the number of photoelectrons generating an output signal of one ADC unit (ADU). It is comprised of the charge to voltage factor at the sense node, external amplification and ADC voltage resolution. Gain can be calculated by measuring X-ray events from a radioactive $^{55}$Fe source \cite{Soman2015}. The decay to $^{55}$Mn produces X-ray photons at 5898 eV and 6490 eV, depositing 1616 and 1778 photoelectrons in Si respectively. Peaks visible at 1060 ADU and 1165 ADU correspond to a system gain of 1.55 $e^{-}$/ADU.

\subsubsection{Read noise}
Read noise has been calculated using dark image subtraction to remove DC offsets. Due to the significant contribution from dark current associated with whole array readouts (825 ms), a 50 row by 376 column ROI is used due to the short readout time of 20 ms. Two images taken in darkness were subtracted and the standard deviation of the difference was divided by $\sqrt 2$ giving a read noise of 4.1 $e^{-}$, agreeing with other reported values for the CIS115 \cite{Soman2015}.

\subsubsection{Linearity}
The photoresponse of the CIS115 was measured by varying the illumination time from a red LED in increments of 500 ms. A window of 200 by 200 pixels was selected when calculating the mean to avoid vignetting. In Figure \ref{fig:Linearity} the measured signal is shown until the deviation from a linear fit reaches 5$\%$. This occurs at 4000 ms, corresponding to a signal of 11235 ADU. The residuals form a polynomial with negative second derivative, due to non-linearity brought on by increasing sense node capacitance at increasing charge levels. Previous work has modelled and corrected for this using a quadratic function \cite{Soman2015}. A quadratic in the form $ADU(t) = a_{1}t+a_{2}t^{2}$ is fit to the data, with the inverse function used to convert the ADU into the time domain (proportional to illuminance), given by equation \ref{quad}. 

The signal vs time ($m_{1}$) where the system gain is measured (1112.5ADU) is then multiplied for gain conversion in e$^{-}$/ADU. After correction is applied, nonlinearity fell to 0.7$\%$. When correcting in the stacked region, the signal ADU are first divided by the FWC increase, $N$, before linearity correction.

\begin{equation}\label{quad}
    \begin{split}
        t(ADU) = \frac{-a_{1}+\sqrt{a_{1}^{2} - 4 a_{2} ADU}}{2a_{2}} \\
        \{ADU\ \epsilon\ R\ |\ 0 \le ADU \le LFWC\}
    \end{split}
\end{equation} 

\begin{figure}[h]
    \centering
    \includegraphics[width=0.5\textwidth]{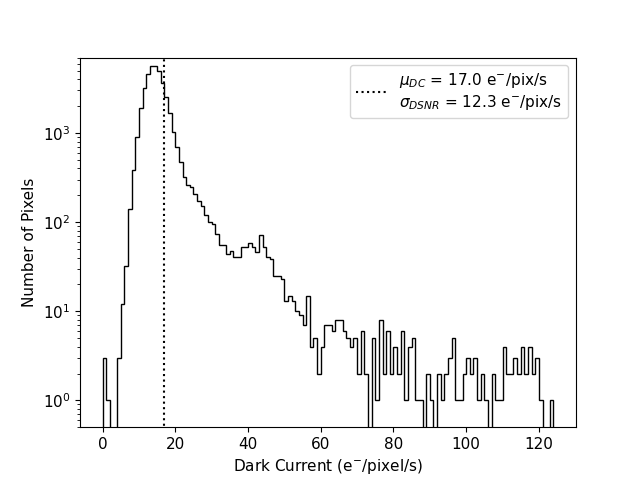}
    \caption{Dark current distribution for 200 by 200 pixels, bin width of 1 e$^{-}$/pixel/s. Dotted line represents the average dark current. The distribution has a component which follows the normal distribution, as well as a tail of pixels with high dark current. Measurements were taken at 294 K.}
    \label{fig:Dark}
\end{figure}
\subsubsection{Dark current}
The dark current of the CIS115 is calculated at room temperature by taking nine integrations with the integration time varying from 0 ms to 4000 ms. This does not include the read time of 10 ms for the 200 by 200 window. Dark current is found for every pixel in the stack and plot as a distribution in Figure \ref{fig:Dark}. Mean dark current was found to be 17.0 e$^{-}$/pix/s with a dark signal nonuniformity of 12.3 e$^{-}$/pix/s. The distribution has a tail of pixels with high dark current. As mentioned previously, dark current is dependent on temperature in an exponential fashion so changes in device temperature could erroneously impact results. All data are reduced with dark frames of equivalent length to help mitigate this.

\begin{figure}[h]
    \centering
    \includegraphics[width=0.5\textwidth]{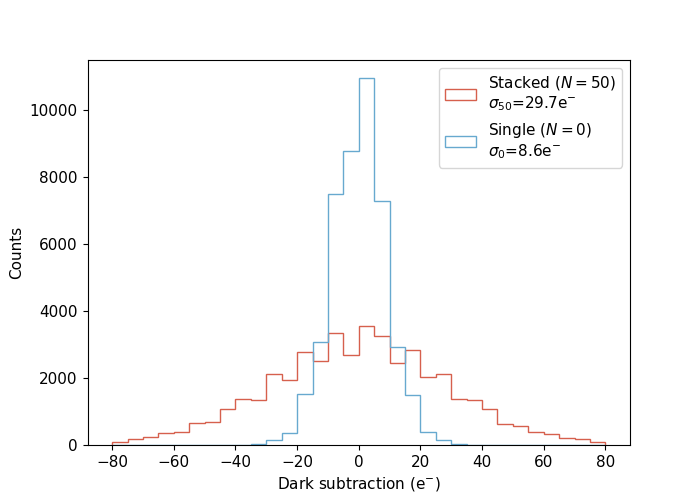}
    \caption{Noise distribution of dark frames for rows in the HDR scheme (red) and standard scheme (blue), bin width of 5 e$^{-}$. Noise increases from 8.6e to 29.7e in the $N=50$ regions due to the read noise adding in quadrature.}
    \label{fig:dark background}
\end{figure}

Figure \ref{fig:dark background} shows the dark frame noise distributions for the standard and HDR schemes during the imaging session in which data for the results section was obtained. For the standard scheme, dark current contributions, $\sqrt{{I}_{dc}t}$, dominate over read noise whereas in the HDR scheme the $N$ reads result in read noise increasing by a factor of $\sqrt{51}$. By using a simple noise model consisting of read and dark current noises (equation \ref{noise}), where $\sigma_{n}$ is the total noise and $t$ the integration time, the dark current is 35.4 e$^{-}$/pix/s, greater than the value found previously at 17.0 e$^{-}$/pix/s. This is due to heating effects, as the PCB and sensor warm up with continual measurements and the sensitive response of dark current to temperature.

\begin{equation}\label{noise}
    \sigma_{n} = \sqrt{(N+1) \times \sigma_{read}^{2} + {I}_{dc}t}
\end{equation}

\subsubsection{Line profile}

The HDR and standard readout schemes are compared using a fixed optics setup to project the Airy disc described previously. $T_{total}$ is 1660 ms for both with $N$ of 50 readouts in the HDR scheme. Row wide line profiles are taken for the first and third Airy reflections allowing comparisons of the schemes in high and low signal regions. Dark frames of equivalent length are subtracted then artefacts corrected for. Centroids of the Airy discs are first located using a 2D Gaussian fitter and window of size 100 pixels. To the nearest pixel the first reflection is located at (1512,951) and third at (315,933). Measurement error is taken as the square root of the signal per pixel as from Poisson statistics. Airy and Gaussian profiles are fit in a window from columns 700 to 1200 as this is where the spatial filter was placed to null Airy rings beyond the second (see Figure \ref{fig:image comparison}). Fit parameters for the Airy disc are intensity, width ($ka\mu/h$, converted to FWHM) and centroid location. For the Gaussian of the fainter third profile these are intensity, width ($\sigma$), centroid location and offset. All are calculated by weighted least squares using the Levenberg-Marquardt algorithm. 

\begin{equation}\label{airy integral}
     \int_{-4}^{4} \frac{1}{8}\left(\frac{J_{1}(r_{p}ka\mu/h)}{r_{p}ka\mu/h} \right)^{2} dr_{p} = 0.998
\end{equation}

Quantitative noise measurement in Figure \ref{fig:PTC high} uses a 9 row 3 column window for noise estimation after light frame subtraction. Due to the non-uniform illumination, shot noise has a spatial variation that increases with window size. The worse case scenario for this is at $r_{p}$=0, due to the average photosignal being 0.2\% lower than I$_{0}$ (equation \ref{airy integral}). At an I$_{0}$ of 1.0 Me$^{-}$ this corresponds to 2.0 ke$^{-}$. The average deviation in this region is 630 e$^{-}$. As the shot noise in this region 1 ke$^{-}$, this effect is in the sub-electron level so is ignored. The window of 27 pixels is chosen to give enough points for a robust noise estimation whilst minimising the aforementioned effect.

\section{Results}
\subsection{High Signal Analysis}

\begin{figure}[h]
    \centering
    \includegraphics[width=0.5\textwidth]{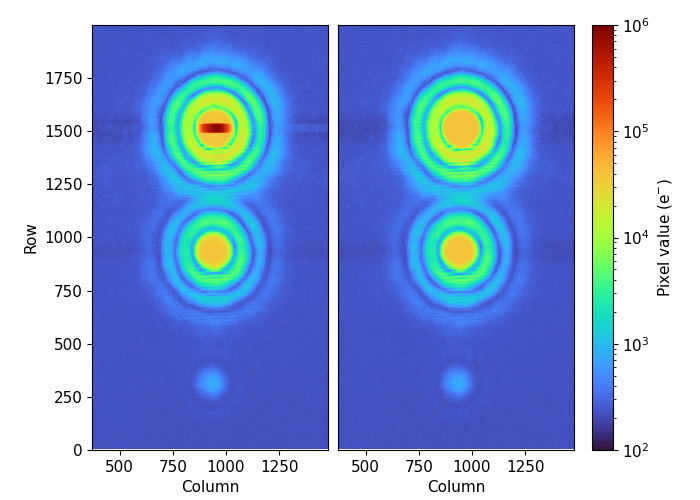}
    \caption{A comparison of the three reflected Airy patterns using the HDR and standard readout schemes. Left is an image of the Airy pattern using the HDR scheme, right is the same pattern with the standard scheme. The majority of pixels in the first order ring of the primary and secondary reflection are saturated at 25 ke$^{-}$. By contrast, pixels up to 937 ke$^{-}$ are recorded in the linear regime for the HDR region.}
    \label{fig:image comparison}
\end{figure}

\begin{figure}[h]
    \centering
    \includegraphics[width=0.5\textwidth]{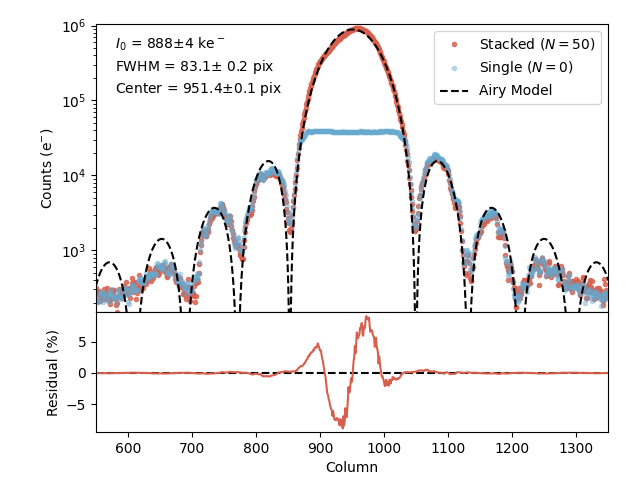}
    \caption{Reduced line profiles of first Airy reflection using HDR and standard readout modes. HDR data is fit to an Airy to model allow for parameter extraction. The standard scheme clips at pixel saturation. Residuals of HDR line profile deviate beyond error bars imposed by Poisson statistics, signifying that deviations in fit are from optical interference and misalignment.}
    \label{fig:reflection 1}
\end{figure}

Maximum photosignal using the HDR scheme in the stacked region, shown by rows containing red pixels in Figure \ref{fig:image comparison}, was measured to be 937 ke$^{-}$ compared to 40 ke$^{-}$ for a single read. This means parts of the projected Airy disc saturated during standard readout but were entirely within the linear regime of the HDR readout scheme. Horizontal line profiles through the centroid of the first reflection are show in Figure \ref{fig:reflection 1}. A least squares fit of an Airy disc to the data gives a FWHM of 83.1$\pm$0.2 pix and I$_{0}$ of 881$\pm$4 ke$^{-}$. Average background signal and standard deviation are measured using columns 0 to 500 of the line profiles to avoid sampling the Airy disc. Values are 236$\pm$44 e$^{-}$ for the HDR and 220$\pm$15 e$^{-}$ for the standard scheme. The HDR profile fits the model better than the standard readout as it remains within the linear region of 1015 ke$^{-}$, whereas the standard readout saturates and clips the data. Higher spatial frequency deviations from the fit of 80 ke$^{-}$ are present, shown by the sinusoidal in the residuals. These can be explained entirely by the optics, namely: interference from the spatial filter, coherence length differences through the plate, optical imperfections, optical alignment and scattered light. The Airy function has repeating minima, light scattering is explained by the photosignal being above 1 ke$^{-}$ at the first minima. This is difficult to mitigate against when the source irradiance is three orders of magnitude greater.

\subsection{Low Signal Analysis}

\begin{figure}[h]
    \centering
    \includegraphics[width=0.5\textwidth]{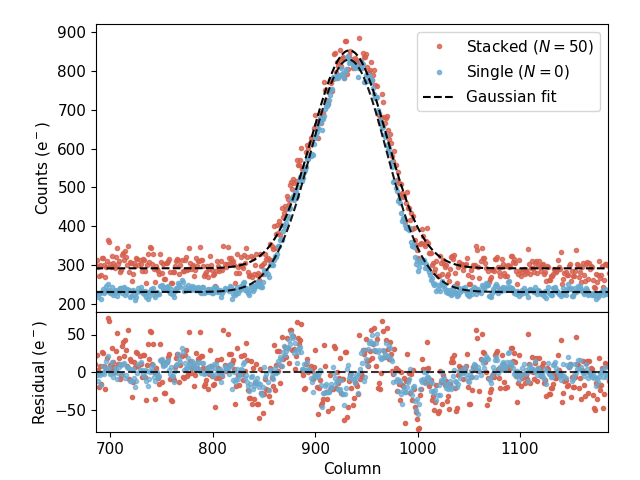}
    \caption{Reduced line profiles of third Airy reflection using HDR and standard readout modes. The HDR scheme shows both higher background noise (34 e$^{-}$ vs 15 e$^{-}$) and background offset (295 e$^{-}$ vs 234 e$^{-}$) compared to the standard readout. Both models are fit to a Gaussian as opposed to an Airy disc as the noise washes out all but the first order.}
    \label{fig:reflection 2}
\end{figure}

The impact of an increased noise floor when measuring small photosignals is demonstrated by taking the line profiles as above for the faintest third Airy disc shown in Figure \ref{fig:reflection 2}. A Gaussian fit with background offset model is employed as orders beyond the first are swamped by the background noise. Fit parameters using weighted least squared for the profiles in range 700 to 1200 columns are given in Table \ref{tab:third params}. The standard scheme has a lower noise floor of 15 e$^{-}$ compared to the HDR scheme at 34 e$^{-}$, visible in the residuals of Figure \ref{fig:reflection 2}. The background signal of the HDR scheme is 292 e$^{-}$, significantly greater than the standard scheme at 231 e$^{-}$ and both background values in the high signal region. This could be due to the previously mentioned signal dependant baseline shift. As a result of the lower noise in the standard scheme, centroiding of the Gaussian FWHM is improved for the standard readout compared to HDR with respective $1\sigma$ uncertainties of 0.2 pixel and 0.3 pixels. Uncertainties in all other model parameters are lower for the standard scheme (see Table \ref{tab:third params}).

\begin{table}[h]
\centering
\begin{tabular}{llll}
\multicolumn{1}{c}{}  & Single readout & \multicolumn{1}{c}{Stacked readout}  \\ \hline
I$_{0}$ (e$^{-}$)      & 597$\pm$3        & 560$\pm$4                \\
$\sigma$ (e$^{-}$)    & 36.9$\pm$0.2      & 36.5$\pm$0.3               \\
Centre (pix)          & 932.9$\pm$0.2     & 933.4$\pm$0.3               \\
Offset (e$^{-}$)       & 231$\pm$1        & 292$\pm$2                 \\
\end{tabular}
\caption{Gaussian fit parameters for Figure \ref{fig:reflection 2}}
\label{tab:third params}
\end{table}

\subsection{SNR Analysis}

\begin{figure}
    \centering
    \includegraphics[width=0.473\textwidth]{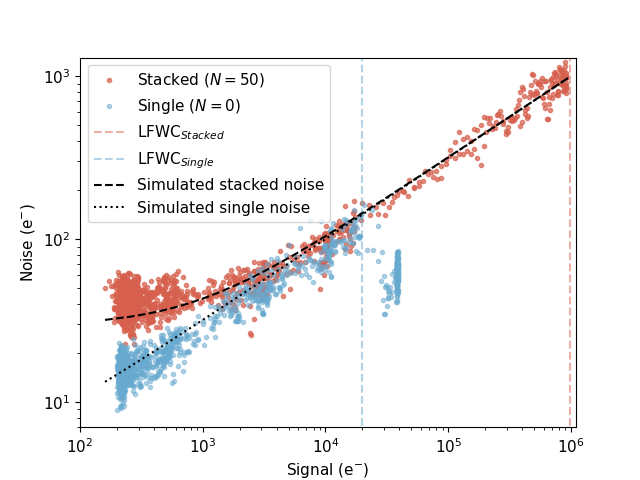}
    \caption{First reflection line profile noise (estimated with 9 $\times$ 3 window) vs window mean for HDR, standard scheme and expected shot noise. The HDR scheme is in a read noise dominated regime at low signals ($<$3 ke$^{-}$) whereas the standard scheme remains in a shot noise regime. Noise for the HDR scheme follows Poisson statistics until peak intensity at 984 ke$^{-}$, whereas the standard readout hits saturation.}
    \label{fig:PTC high}
\end{figure}

Figure \ref{fig:PTC high} shows the distribution of noise against signal for the HDR (red) and standard (blue) schemes as in Figure \ref{fig:reflection 1}. Simulated noise performance is overlaid as calculated in Figure \ref{fig:PSNR} by noise = $\sqrt{(N+1)\times4.1^{2} + S}\ (e^{-})$, where $S$ is the signal in $e^{-}$. For both schemes the measured noise bisects the simulated noise until the LFWC is reached. Noise is significantly higher in the stacked scheme at low signals until approximately 3 ke$^{-}$ where shot noise makes up 90\% of the total noise (see Figure \ref{fig:PSNR}). This could be taken as the selection criteria for where to position rows read once. Signals beyond the LFWC of the standard scheme create an island of data points that deviate significantly below the expected signal shot noise due to sensor saturation. The HDR scheme extends beyond this with data bisecting the simulated noise to the LFWC of 984 ke$^{-}$, validating the linearity preserving behaviour of this method.

\subsection{Dynamic Range Analysis}

For a given imaging target, HDR readout rows can be interlaced with a standard readout, with the benefit of lower baseline noise in low signal regions (Figure \ref{fig:reflection 2}) and arbitrarily high FWC in high signal regions (Figure \ref{fig:reflection 1}). For calculations of dynamic range, the read noise is used as the noise floor. This is because of experimental limitations, as dark current noise and stray light shot noise were dominant. Active cooling and the use of a flat illumination source has been used to suppress both sources in similar CIS devices \cite{Wang2020}.

Compared to the standard readout scheme, interleaving readouts in high signal regions results in an increase in dynamic range from 73.6 dB to 107.6 dB. Minimum exposure time is increased, depending on the number of additional row reads. Critically, the extended dynamic range stays in the linear regime of the CIS, with the ability to correct for nonlinearity. In this work, readout order was predefined for a known scene, however a shortened sub-exposure and readout order optimisation algorithm could dynamically adapt the row readout order to maximise the DR. 

\begin{table*}
\centering
\resizebox{\textwidth}{!}{%
\begin{tabular}{lllllll}
\multicolumn{1}{c}{}  & This paper & \cite{Peizerat2015} & \cite{Hirsch2019} & \cite{simsCMOSVisibleImage2018} & \multicolumn{1}{c}{\cite{dupontDualcoreHighlyProgrammable2016}}  \\ \hline
Method & Row-wise coded exposure & Block coded exposure & Self reset & Non-destructive readout  & Row alternating exposure    \\
Pitch ($\mu$m) &7.0 & 5.0 & 20.0 & 16.0 & 10.0 \\
Array size  & 2000$\times$1504 & 512$\times$800 & 16$\times$16 & 2048$\times$2048 & 2800$\times$1088 \\
Read noise (e$^{-}$) & 4.1  &  -  & - & 9.8 (1 NDR) & 2.6            \\
Linear FWC (e$^{-}$)   & 19,680  & - & - & 20,000 & 100,000      \\
DR (dB)     & 107.6 (50 samples)  & 120 & 121 & 144 (200 NDR) & 120                 \\
Relative fill factor   & High & High & Low & High & High              \\
DR fidelity & Per row & Per 32$\times$32 block & Per pixel & Per pixel & Per alternating row \\
PSNR increase & Yes & No & Yes  & No & Limited (from dual gain) \\
\end{tabular}
}
\caption{Comparative Summary of HDR Extension Demonstrations}
\label{tab:DR}
\end{table*}

\subsection{Applications}

A future implementation of this technique is for actively stabilised astronomical spectroscopy, where it is desirable for both calibration spectral lines and science (observation performed) spectral lines to have the same optical path to the focal plane. Stability is achieved through the continuous generation of corrections to the focal plane from the centroids of calibration spectral lines \cite{Jones2021}. These continuous small corrections are possible because the brighter calibration signal is read out frequently whilst the fainter stellar science signal builds up on adjacent regions of the detector which are read out at a much lower rate. Emission lines from traditional calibration sources can also vary in strength. Maximising uniformity in intensity and therefore effective spatial uniformity in all regions of interest leads to the potential for more precise corrections to the focal plane \cite{Li2008}. This in turn enables more precise scientific measurements. Although laser-comb technology has resolved these non-uniformity issues, they are much more expensive to build and maintain and do not overcome the issue of focal plane stability.

In principle, there are also a variety of other applications where multi-cadence ROIs might be important. As with the case of astronomical spectroscopy, telescope time on the target is a very valuable commodity. In most types of astronomical imaging it is desirable to obtain the longest exposure and highest dynamic range possible before taking the penalty of readout noise. For example, when trying to discriminate planets in orbit around stars there is typically a contrast ratio in excess of a million and so there are many different techniques used to reduce the contrast ratio \cite{2010exop.book..111T} and make signal detection more manageable for detectors.

\section{Conclusion}

An implementation of row-wise coded exposure using a CIS115 increased the dynamic range from 73.6dB to 107.6dB and PSNR from 42.9 dB to 59.9 dB. The utility of this operation is demonstrated whereby the linear FWC in high signal 'calibration' regions was extended 50 times with an interlaced readout order, whilst the noise floor remained unchanged with a single readout in low signal 'science' regions. The DR can be extended arbitrarily high by reading out a row multiple times. This scheme resulted in an increase of centroiding precision in low signal regions due to the single read, whilst extending the linear FWC via multiple readouts. Row readout order needs to be pre-defined making it more suitable for well-known scenes with little temporal variation. Further work would involve a working demonstration in an echelle spectrograph and the exploration of simultaneous ROI imaging at varying sub-exposure durations.

A comparative summary is given in Table \ref{tab:DR}. Although this paper achieves the smallest absolute DR compared to other works in this table, the extension range is not hardware based so can be increased to a user defined level at the cost of increased minimum integration time. By increasing the number of reset and read samples from 50 to 200, the DR extension demonstrated in this paper would increase to 120 dB. This is achieved on 7.0 $\mu$m pitch whilst preserving linearity. The only other technique to extend PSNR continuously is the self reset pixel, which comes at the cost of low fill factor. In this work the DR is increased on a per row basis; independent column addressability would extend the method to DR increase per pixel \cite{liuEfficientSpaceTimeSampling2014a}.

\section*{Acknowledgement}
We acknowledge the support of Teledyne e2v in providing the CIS115. We add thanks to Andrew Pike, Douglas Jordan and Paul Jerram from Teledyne e2v for their support.

For the purpose of open access, the authors have applied a Creative Commons Attribution (CC BY) licence to any Author Accepted Manuscript version arising.

Data underlying the results presented in this paper can be requested by emailing t.wocial@herts.ac.uk

% Bibliography info: in-text citations are done using biblatex. Due to not being able to format to the exact IEEE requirements, the bibliography is manually entered (below) using the IEEE template. 

% Use command \printbibliography below to print the unformatted in-text citations using biblatex. This HAS TO MATCH the order (not formatting) of the manual bibliography. 

\printbibliography 

\end{document}